\documentclass[12pt]{article}
\usepackage{amssymb,amsmath,amsthm,amsfonts,amscd}
 \usepackage{graphicx}
\newcommand\be{\begin{equation}}
\newcommand\ee{\end{equation}}
\newcommand\bea{\begin{eqnarray}}
\newcommand\eea{\end{eqnarray}}
\newcommand\ket[1]{|#1\rangle}
\newcommand\bra[1]{\langle #1|}

\newcommand{\fatalpha}{{\bf \alpha \kern -0.44em \alpha}}
\newcommand{\fatsigma}{{\bf \sigma \kern -0.54em \sigma}}
\newcommand{\tpchi}{{\bf D \kern -0.35em D}}
\newcommand{\llambda}{{\bf \lambda \kern -0.45em \lambda}}

\renewcommand{\theequation}{\arabic{equation}}
\renewcommand{\theequation}{\thesection.\arabic{equation}}
\bibliography{plain}
\title{\bf Detecting three-qubit bound MUB diagonal entangled  states via
Nonlinear optimal entanglement witnesses } \vspace{20mm}
\author{ M. A. Jafarizadeh$^{a,b,c}$
 \thanks{E-mail:jafarizadeh@tabrizu.ac.ir}, M. Mahdian$^{a}$
 \thanks{E-mail:Mahdian@tabrizu.ac.ir},
 A. Heshmati$^{a}$ \thanks{E-mail:Heshmati@tabrizu.ac.ir}, K. Aghayari$^{a}$
 \thanks{E-mail:Aghayari@tabrizu.ac.ir}\\
$^a${\small Department of Theoretical Physics and Astrophysics,
University of Tabriz, Tabriz 51664, Iran.}  \\ $^b${\small
Institute for Studies in Theoretical Physics and Mathematics,
Tehran 19395-1795, Iran.}\\$^c${\small Research Institute for
Fundamental Sciences, Tabriz 51664, Iran.}} \pagebreak

\vspace{20mm}

\begin{document}
\maketitle \vspace{15mm}
\newpage
\begin{abstract}
One of the important approaches to detect quantum entanglement is
using linear entanglement witnesses (\emph{EW}s). In this paper, by
determining the envelope of the boundary hyper-planes defined by a
family of linear \emph{EW}s, a set of powerful nonlinear optimal \emph{EW}s is
manipulated. These \emph{EW}s enable us to detect some three qubits bound
\emph{MUB} (mutually unbiased bases) diagonal entangled states, i.e., the
\emph{PPT} (positive partial transpose) entangled states. Also, in some
particular cases, the introduced nonlinear optimal \emph{EW}s are
powerful enough to separate the bound entangled regions from the
separable ones. Finally, we present numerical examples to
demonstrate the practical accessibility of this approach.

\end{abstract}

\newpage
\section{Introduction}
  In the recent years it became clear that quantum entanglement
\cite{Einstein} is one of the most important resources in the
rapidly expanding field of quantum information processing, with
remarkable applications such as quantum parallelism
\cite{Deutsch}, quantum cryptography \cite{Ekert}, quantum
teleportation \cite{Bennett,Bouwmeester}, quantum dense coding
\cite{Wiesner,Mattle} and reduction of communication complexity
\cite{Cleve}. The above ideas are based on the fact that quantum
entanglement, in particular, the occasionally occurrence of
entangled states produce nonclassical phenomena. Therefore,
specifying that a particular quantum state is entangled or
separable is important because if the quantum state be separable
then it's statistic properties can be explained entirely by
classical statistics.\par In this paper, we will deal with three
qubit systems with $2^3$-dimensional Hilbert space
${\cal{H}}_{2}\otimes {\cal{H}}_{2}\otimes {\cal{H}}_{2}$,
(${\cal{H}}_{d}$ denotes the Hilbert space with dimension
\emph{d}). A density matrix $\rho$ on this Hilbert space, is
called fully separable if it can be written as a convex
combination of pure product states as follows
\begin{equation}\label{fullsep}
    \rho=\sum_{i} p_{i} | \alpha_{i}^{(1)} \rangle \langle \alpha_{i}^{(1)} |\otimes
    | \alpha_{i}^{(2)} \rangle \langle \alpha_{i}^{(2)}
    |\otimes | \alpha_{i}^{(3 )} \rangle \langle \alpha_{i}^{(3)}| ,
\end{equation}
where $|\alpha_{i}^{(j)}\rangle$ are  arbitrary but normalized
vectors lie in ${\cal{H}}_{2}$, and $p_{i}\geq0$ satisfy
$\sum_{i}p_{i}=1$ (hereafter we will refer to fully separable
states as separable ones). The first and most widely used related
criterion for distinguishing entangled states from separable ones,
is the Positive Partial Transpose (\emph{PPT}) criterion,
introduced by Peres \cite{peres}. Furthermore, the necessary and
sufficient condition for separability in ${\cal{H}}_{2}\otimes
{\cal{H}}_{2}$ and
 ${\cal{H}}_{2}\otimes
{\cal{H}}_{3}$ was shown by  Horodecki in Ref. \cite{horodecki1},
which was based on a previous work by Woronowicz
\cite{woronowicz}. Partial transpose means transposition with
respect to one of the subsystems. For a quantum state
$\rho_{_{AB}}$ with matrix entries $\rho_{ij}^{mn}=\langle
ij|\rho_{AB}|mn\rangle$,
 the partial transposition with respect to the subsystem $B$, denoted by $\rho_{AB}^{T_B}$, is defined by $$(\rho_{ij}^{mn})^{T_B}=\rho_{in}^{mj}.$$
 However, as it was shown in Ref. \cite{horodecki2}, in higher
dimensions, there are \emph{PPT} states that are nonetheless
entangled. These states are called \emph{PPT} entangled states
(\emph{PPTES}) or bound entangled states because they possess the
peculiar property that no entanglement can be distilled from them
by local operations \cite{horodecki3}. Another approach to
distinguish separable states from entangled ones involves the so
called entanglement witness (\emph{EW}) \cite{terhal}. An
\emph{EW} for a given entangled state $\rho$ is an observable
$\emph{W}$ whose expectation value over all separable states is
nonnegative, but strictly negative on $\rho$. There is a
correspondence between EWs and linear positive (but not completely
positive) maps via Jamiolkowski isomorphism \cite{jamiolkowski}.
As an example the partial transposition is a positive map
(PM).\par In this work, we consider those density matrices which
are written as a linear combination of maximally commuting
observables taken from the set of tensor products $A^i\otimes
B^j\otimes C^k$, where
$A,B,C\in\{I_2,\sigma_x,\sigma_y,\sigma_z\}$ and $i,j,k\in
\{0,1\}$ ($\sigma_x$, $\sigma_y$ and $\sigma_z$ are usual Pauli
matrices). We will see later on that common eigenvectros of these
observables form mutually unbiased bases
(\emph{MUB})\cite{Wootters1} and so we will refer to a set of such
observables as set of \emph{MUB} observables, for instance by
using the notation $\sigma_i\sigma_j\sigma_k\equiv
\sigma_i\otimes\sigma_j\otimes\sigma_k$, the set $\{I I I
,\sigma_{z}\sigma_{z}I, \sigma_{z} I \sigma_{z},
I\sigma_{z}\sigma_{z}
,\sigma_{x}\sigma_{x}\sigma_{x},\sigma_{x}\sigma_{y}\sigma_{y},
\sigma_{y} \sigma_{x}\sigma_{y},\sigma_{y} \sigma_{y}\sigma_{x}\}$
is a set of \emph{MUB} observables. In fact, we consider
tripartite \emph{MUB} diagonal density matrices which are written
in terms of \emph{MUB} observables in a diagonal form. Then, we
impose the \emph{PPT} conditions (positivity of partial
transposition with respect to all subsystems) to these density
matrices and refer to the region of those density matrices which
satisfy all of the obtained \emph{PPT} conditions as ``feasible
region" (see Fig.1 and Fig.2 for example). In this way, we see
that partial transposition plays an important role because in this
type of density matrices, conditions obtained from positivity of
partial transpositions are linear and feasible regions are
completely contained in polygons; this allows us to investigate
the separability or entanglement of the density matrices. In order
to distinguish PPT entangled states (\emph{PPTES}) from separable
ones we construct some linear and nonlinear \emph{EW}s. Namely, we
consider density matrices that their common eigenvectors are
maximally entangled states (\emph{GHZ}-states) and construct an
\emph{EW} that detects such density matrices. Finally, we consider
three categories relevant to some special choices of the
parameters of density matrices, and by using the linear EWs we
distinguish the region of \emph{PPTES} and separable states
completely. In other words, for density matrices contained in one
of these three categories, we show that if the introduced linear
EWs can not detect their entanglement, then they are necessarily
separable.\par We have also provided some numerical evidence
suggesting that \emph{PPTES} of three qubits can be detected by
using the nonlinear EWs.

The paper is organized as follows: In section 2, we introduce the
\emph{MUB}-$(\emph{zzz})_G$ diagonal density matrices and consider
the corresponding \emph{PPT} conditions and feasible region.
Section 3 is devoted to definition of an \emph{EW} and
construction of optimal linear \emph{EW}s. In section 4, we obtain
an envelope of family of linear \emph{EWs} and construct some
nonlinear \emph{EW}s. Section 5 is devoted to classification and
detection of \emph{PPTES} for \emph{MUB} diagonal density matrices
in three categories. In section 6, we discuss some numerical
analysis for evaluating the feasible region and the region of
\emph{PPTES}. The paper is ended with a brief conclusion together
with two appendices.
\section{\emph{MUB} diagonal density matrices}
In this section we introduce the so called \emph{MUB} diagonal
density matrices. The basic notions and definitions of \emph{MUB}
states relevant to our study are given in the Appendix $I$.
\subsection{\emph{MUB}-$(\emph{zzz})_G$ diagonal density matrices}\par
In this subsection we introduce the \emph{MUB}-$(\emph{zzz})_G$
diagonal density matrices which are considered through the paper.
A \emph{MUB}-$(\emph{zzz})_G$ diagonal density matrix for three
qubits is defined as
\\ \be \label{ro r} \rho=\sum_{i=1}^{8} p_{i}|\psi_{i}\rangle\langle\psi_{i}|
\ , \ 0\leq p_{i} \leq 1 \ , \ \sum_{i=1}^{8}p_{i}=1 ,\ee
 where three qubit \emph{GHZ} states $|\psi_{i}\rangle$ for $i=1,2,...,8$ are given by
$$|\psi_{1}\rangle=\frac{1}{\sqrt{2}}[ |000\rangle + |111\rangle ]  ,\ |\psi_{2}\rangle=\frac{1}{\sqrt{2}}[ |000\rangle - |111\rangle ] ,$$
$$|\psi_{3}\rangle=\frac{1}{\sqrt{2}}[ |001\rangle + |110\rangle ]  ,\ |\psi_{4}\rangle=\frac{1}{\sqrt{2}}[ |001\rangle - |110\rangle ] ,$$
$$|\psi_{5}\rangle=\frac{1}{\sqrt{2}}[ |010\rangle + |101\rangle ]  ,\ |\psi_{6}\rangle=\frac{1}{\sqrt{2}}[ |010\rangle - |101\rangle ] ,$$
\be\label{bell}|\psi_{7}\rangle=\frac{1}{\sqrt{2}}[ |011\rangle +
|100\rangle ] ,\ |\psi_{8}\rangle=\frac{1}{\sqrt{2}}[ |011\rangle
- |100\rangle ].\ee Then, by using the \emph{MUB} states in line
$6$ of Table $I$ given in Appendix $A$, the density matrix $\rho$
can be rewritten as follows \be\label{bell ghz}\rho=\frac{1}{8}[ I
I I + r_{1} \sigma_{z}\sigma_{z}I+ r_{2} \sigma_{z} I \sigma_{z}+
r_{3} I\sigma_{z}\sigma_{z} + r_{4}\sigma_{x}\sigma_{x}\sigma_{x}+
r_{5}\sigma_{x}\sigma_{y}\sigma_{y}+ r_{6} \sigma_{y}
\sigma_{x}\sigma_{y}+ r_{7} \sigma_{y} \sigma_{y}\sigma_{x} ], \ee
where
$$r_{1}= +p_{1}+p_{2}+p_{3}+p_{4}-p_{5}-p_{6}-p_{7}-p_{8},$$
$$r_{2}= +p_{1}+p_{2}-p_{3}-p_{4}+p_{5}+p_{6}-p_{7}-p_{8},$$
$$r_{3}= +p_{1}+p_{2}-p_{3}-p_{4}-p_{5}-p_{6}+p_{7}+p_{8},$$
$$r_{4}= +p_{1}-p_{2}+p_{3}-p_{4}+p_{5}-p_{6}+p_{7}-p_{8},$$
$$r_{5}=-p_{1}+p_{2}+p_{3}-p_{4}+p_{5}-p_{6}-p_{7}+p_{8},$$
$$r_{6}=-p_{1}+p_{2}+p_{3}-p_{4}-p_{5}+p_{6}+p_{7}-p_{8},$$
\be\label{rp}r_{7}=-p_{1}+p_{2}-p_{3}+p_{4}+p_{5}-p_{6}+p_{7}-p_{8}
.\ee It should be noticed that the \emph{MUB} states of any line
of Table $I$ except for the states in the first three lines which
are associated with separable states, can define a \emph{MUB}
diagonal density matrix, similarly. In the next subsection we
impose the \emph{PPT } conditions to the density matrix (\ref{bell
ghz}) in order to obtain the corresponding feasible region (region
of those \emph{MUB} diagonal density matrices which satisfy all of
the \emph{PPT} conditions).
\subsection{ Feasible regions }
Here we are concerned with \emph{MUB} diagonal density matrices of
type (\ref{bell ghz}) and by imposing the conditions obtained from
positivity of partial transpositions with respect to each qubit, we
obtain the so called feasible region. For these particular density
matrices, the positivity of partial transpositions gives linear
constraints on the parameters $p_i$ for $i=1,2,...,8$. In order to
obtain the feasible region for density matrix (\ref{bell ghz}),
first we group the conditions obtained from positivity of partial
transpositions with respect to each subsystem (each qubit) in six
partitions $(p_3,p_4,p_5,p_6)$, $(p_1,p_2,p_7,p_8)$,
$(p_1,p_2,p_5,p_6)$, $(p_3,p_4,p_7,p_8)$, $(p_1,p_2,p_3,p_4)$ and
$(p_5,p_6,p_7,p_8)$ as follows:
\par The positivity of partial transposition with respect to the first qubit gives the following constraints:
\be\label{p1}(p_3,p_4,p_5,p_6)\equiv\left\{\begin{array}{c}p_3 + p_4 + p_5 - p_6\geq0\\
p_3 + p_4 - p_5 + p_6\geq0\\p_3 - p_4 + p_5 + p_6\geq0\\-p_3 + p_4 + p_5 + p_6\geq0\\
\end{array}\right.\ee

\be\label{p2}(p_1,p_2,p_7,p_8)\equiv\left\{\begin{array}{c}p_1 + p_2 + p_7 - p_8\geq0\\
p_1 + p_2 - p_7 + p_8\geq0\\p_1 - p_2 + p_7 + p_8\geq0\\ -p_1 + p_2 + p_7 + p_8\geq0\\
\end{array}\right.\ee
The positivity of partial transposition with respect to the second
qubit gives:
\be\label{p3}(p_1,p_2,p_5,p_6)\equiv\left\{\begin{array}{c}p_1 + p_2 + p_5 - p_6\geq0\\
p_1 + p_2 - p_5 + p_6\geq0\\p_1 - p_2 + p_5 + p_6\geq0\\-p_1 + p_2
+ p_5 +
p_6\geq0\\
\end{array}\right.\ee
\be\label{p4}(p_3,p_4,p_7,p_8)\equiv\left\{\begin{array}{c}p_3 + p_4 + p_7 - p_8\geq0\\
p_3 + p_4 - p_7 + p_8\geq0\\p_3 - p_4 + p_7 + p_8\geq0\\-p_3 + p_4
+
p_7 + p_8\geq0\\
\end{array}\right.\ee
The positivity of partial transposition with respect to the third
qubit gives:
\be\label{p5}(p_1,p_2,p_3,p_4)\equiv\left\{\begin{array}{c}p_1 + p_2 + p_3 - p_4\geq0\\
p_1 + p_2 - p_3 + p_4\geq0\\p_1 - p_2 + p_3 + p_4\geq0\\-p_1 + p_2 + p_3 + p_4\geq0\\
\end{array}\right.\ee
\be\label{p6}(p_5,p_6,p_7,p_8)\equiv\left\{\begin{array}{c}p_5 + p_6 + p_7 - p_8\geq0\\
p_5 + p_6 - p_7 + p_8\geq0\\p_5 - p_6 + p_7 + p_8\geq0\\-p_5 + p_6 + p_7 + p_8\geq0\\
\end{array}\right.\ee
The region of those density matrices of type (\ref{bell ghz})
which satisfy the above $24$ constraints, is the feasible region.
In order to specify the new perspective from this feasible region,
we consider the parameters $p_i$ in four pairs $(p_1,p_2)$,
$(p_3,p_4)$, $(p_5,p_6)$ and $(p_7,p_8)$.\par Now if we choose one
of the pairs, say $(p_1,p_2)$, then we can specify the projection
of the feasible region to $(p_1,p_2)$ plane with the following
three inequalities ( the last inequalities of (\ref{p2}),
(\ref{p3}) and (\ref{p5}), respectively\emph{PPT} conditions)
$$\left\{\begin{array}{c}p_1\leq p_2+p_7+p_8\\p_1\leq p_2+p_5+p_6\\p_1\leq p_2+p_3+p_4\\
\end{array}\right.$$
By adding right hand side and left hand side of the above
inequalities and using the equality $\sum_{i=1}^{8}p_{i}=1$, we
get the following inequality \be\label{eq1}4p_1-2p_2\leq1.\ee
Similarly, one can obtain the inequality  \be\label{eq1'}
4p_2-2p_1\leq1,\ee by exchanging $1$ and $2$ in the above steps.
For an illustration see Fig. $1$. It should be noticed that if we
chose any other pair from $(p_3,p_4)$, $(p_5,p_6)$ and
$(p_7,p_8)$ instead of the pair $(p_1,p_2)$, we would obtain the similar inequalities as in (\ref{eq1}) and (\ref{eq1'}) for each pair. \\
According to Fig. 1, since the vertex points $(\frac{1}{2},
\frac{1}{2})$, $(\frac{1}{4} , 0)$, $(0 , \frac{1}{4})$ and (0,0)
satisfy the criterion (\ref{eq1}), all of the points inside the
shape will fulfill the \emph{PPT} conditions (this is due to the
fact that the feasible region is a convex region).\par We can find
another new projection of the feasible region in the $(p_1 ,p_3)$
plane, concerning the following inequalities
\be\label{eq2}\left\{\begin{array}{c}p_1\leq p_2+p_5+p_6\\
p_3\leq p_4+p_7+p_8\\
\end{array}\right.\Rightarrow p_1+p_3\leq \frac{1}{2}.\ee
This region is illustrated in Fig. $2$; therefore we have
presented a projection of the spatial shape in a two-dimensional
space.
\subsection{A special case of feasible region}
Here, we discuss a special case of feasible region which will be
appeared in subsection 5.3 as a region of bound entangled MUB
diagonal density matrices (PPT entangled states). To this aim, we
consider the line $p_1+p_3=\frac{1}{2}$ of the feasible region
(\ref{eq2}) in the $(p_1,p_3)$ plane (see Fig. $2$).

First we take the feasible region for the $(p_3,p_4)$ plane (see
Fig.3). If, we consider the following parametric line equation
\be\label{eq5}p_3=\alpha p_4+\frac{1}{4},\ee then (according to
equations similar to (\ref{eq1}) and (\ref{eq1'}) for the pair
$(p_3,p_4)$), we obtain \be\label{x}4p_4-2p_3=1\Rightarrow
p_4=\frac{p_3}{2}+\frac{1}{4}\ee By substituting (\ref{x}) in
(\ref{eq5}) and using the fact that $0\leq p_3\leq 1/2$ (see Eq.
(\ref{eq2})), one can obtain
$$p_3=\frac{\alpha p_3}{2}+\frac{\alpha+1}{4}\ \ \Rightarrow \ \ \
0\leq p_3=\frac{\alpha+1}{4-2\alpha}\leq\frac{1}{2} ,$$ so we get
$-1\leq\alpha\leq \frac{1}{2}$. According to the boundary
condition $p_1+p_3=\frac{1}{2}$ and (\ref{eq5}) we obtain:
\begin{equation}\label{eqx}
p_1=-\alpha p_4+\frac{1}{4} .
\end{equation}

Now by considering the PPT conditions
\be\label{in1}\left\{\begin{array}{c}-p_3+p_4+p_5+p_6\geq0\\
-p_3+p_4+p_7+p_8\geq0\\
\end{array}\right. ,\ee
from equations (\ref{p1}) and (\ref{p4}), and adding the sides of
the them, we obtain \be\label{eq11}(1-2\alpha)p_4-p_2\geq0.\ee
Also from the PPT conditions
\be\label{in2}\left\{\begin{array}{c}-p_1+p_2+p_5+p_6\geq0\\
-p_1+p_2+p_7+p_8\geq0\\
\end{array}\right. ,\ee given in (\ref{p2}) and (\ref{p3}), we get
\be\label{eq12}-(1-2\alpha)p_4+p_2\geq0.\ee Then, from
(\ref{eq11}) and (\ref{eq12}) we conclude that
$p_2=(1-2\alpha)p_4$. Then, by using (\ref{eq5}) and (\ref{eqx})
one can easily conclude the following equations
$$p_3-p_4=(\alpha-1)p_4+\frac{1}{4}$$
$$p_1-p_2=-\alpha p_4+\frac{1}{4}-(1-2\alpha)p_4=(\alpha-1)p_4+\frac{1}{4},$$ which indicate that $p_3-p_4=p_1-p_2$. On the other hand, from the
inequalities (\ref{eq11}) and (\ref{eq12}), one can deduce that
the left hand sides of the inequalities (\ref{in1}) and
(\ref{in2}) must be equal to zero. Therefore, for PPT density
matrices with positive $p_i$'s, we obtain
$$p_5+p_6=p_7+p_8=p_3-p_4=(\alpha-1)p_4+\frac{1}{4}\geq0 \ \ \Rightarrow
\ \ p_4\leq\frac{1}{4(1-\alpha)}$$ furthermore we obtain $p_3\geq
p_4.$ Clearly, the PPT conditions (\ref{p5}) and (\ref{p6}) are
satisfied by using the relations $p_1+p_4=p_2+p_3$ and
$p_5+p_6=p_7+p_8$. The bound entanglement or separability of
density matrices belonging to this special case of feasible region
will be discussed in subsection $5.3$ (as third category). For an
illustration see Fig.3.
\subsection{MUB-$(zzz)_G$ diagonal density matrices for which PPT conditions are necessary and
sufficient for separability} In this section we consider the
family of MUB-$(zzz)_G$ diagonal density matrices where the PPT
criterions are necessary and sufficient for their separability.
\subsubsection{Case (1)}
In this case, we will put one of the pairs $(p_1,p_2),(p_3,p_4),
(p_5,p_6)$ and $(p_7,p_8)$ equal to $(0,0)$, then we will see that
the PPT conditions given in (\ref{p1})-(\ref{p6}) are necessary
and sufficient for separability of MUB-$(zzz)_G$ diagonal density
matrices. For example, if we choose
$$p_1=p_2=0,$$ then, by using the PPT conditions, we obtain $$p_{3}=p_{4} ,\;\ p_{5}=p_{6}
,\;\ p_{7}=p_{8}.$$Then, the density matrices satisfying these
conditions can be written as \be\label{ro}\rho=\frac{1}{8}[ I I I
+ r_{1} \sigma_{z}\sigma_{z} I+ r_{2} \sigma_{z} I \sigma_{z}+
r_{3} I \sigma_{z}\sigma_{z}  ] .\ee The density matrix $\rho$ in
(\ref{ro}) is separable, since by using (\ref{rp}) we can rewrite
$\rho$ as
$$\rho=\frac{1}{4}\{p_3(III+ \sigma_{z}\sigma_{z}
I- \sigma_{z} I \sigma_{z}-  I \sigma_{z}\sigma_{z}) $$
$$+p_5(III+
\sigma_{z} I \sigma_{z} - \sigma_{z} \sigma_{z} I-  I
\sigma_{z}\sigma_{z})+p_7(III+ I \sigma_{z}\sigma_{z} I-
\sigma_{z} I \sigma_{z}- \sigma_{z}\sigma_{z} I) \}=$$
$$
p_3(\ket{\psi_3}\bra{\psi_3}+\ket{\psi_4}\bra{\psi_4})
+p_5(\ket{\psi_5}\bra{\psi_5}+\ket{\psi_6}\bra{\psi_6})+p_7(\ket{\psi_7}\bra{\psi_7}+\ket{\psi_8}\bra{\psi_8})
,$$ which is clearly a separable state, since it is a convex
combination of projection operators.
\subsubsection{Case (2)}
In this case, we choose $p_i$'s in each pair except for one of
them to be equal, then we show that the \emph{PPT} conditions
(\ref{p1})-(\ref{p6}) are necessary and sufficient for
separability of \emph{MUB}-$(zzz)_G$ diagonal density matrices.
For example, we consider $$p_{1}\neq p_2 ,\;\  p_3=p_4 ,\;\
p_5=p_6 ,\;\ p_7=p_8 .$$ Then, we can write the density matrix
(\ref{ro r}) as follows
$$\rho = (\frac{p_1 +p_2}{2}) ( |\psi_1 \rangle \langle \psi_1 |+ |\psi_2 \rangle \langle \psi_2 | )+(\frac{p_1-p_2}{2})( |\psi_1 \rangle \langle
\psi_1 |-  |\psi_2 \rangle \langle \psi_2 | )+
 p_3 (|\psi_3 \rangle \langle \psi_3
|+ |\psi_4 \rangle \langle \psi_4 | )+$$ \be\label{rox}p_5
(|\psi_5 \rangle \langle \psi_5|+ |\psi_6 \rangle \langle \psi_6 |
)+ p_7 (|\psi_7 \rangle \langle \psi_7 |+|\psi_8 \rangle \langle
\psi_8 |).\ee
We assume that $p_3<p_5<p_7 $ and  $p_2<p_1 $ (the other cases give the same results as those which is obtained by this assumption in the following).\\
By substituting $p_3=p_4$ in \emph{PPT} conditions ($\ref{p5}$),
we obtain $p_1\leq p_2+2p_3$, so that we can write $p_1=p_2+2p_3-2
\epsilon_1 $ $(p_1>p_2 \Rightarrow 0 \leq \epsilon_1 \leq p_3)$.
Also, from the assumptions $p_3<p_5$ and $p_3<p_7$, one can write
$p_5=p_3+\epsilon_5 $ and $p_7=p_3+\epsilon_7$, respectively. By
substituting these values of $p_i$'s in the density matrix
(\ref{rox}) and using the resolution of identity
$\sum_{i=1}^8|\psi_i \rangle \langle \psi_i |=III$, one can write
$$\rho=\epsilon_1 III +(p_2-\epsilon_1)(|\psi_1 \rangle \langle \psi_1 |+|\psi_2 \rangle \langle \psi_2 |)+(p_3-\epsilon_1)(III +|\psi_1 \rangle \langle \psi_1 |-|\psi_2 \rangle \langle \psi_2 |)$$
\be\label{eqxx}+\epsilon_5(|\psi_5 \rangle \langle \psi_5
|+|\psi_6 \rangle \langle \psi_6 |)+\epsilon_7(|\psi_7 \rangle
\langle \psi_7 |+|\psi_8 \rangle \langle \psi_8 |).\ee
 Then, from the fact that $(III\pm |\psi_1 \rangle \langle \psi_1
|-|\psi_2 \rangle \langle \psi_2 |)$ are separable states, one can
see that for $\epsilon_1 <p_2$ in (\ref{eqxx}), the density matrix
$\rho$ is separable (since it is written as a convex combination
of product states). For $\epsilon_1
>p_2$, we can write $\rho$ as follows
$$\rho=p_2 III +(\epsilon_1 -p_2)(III-|\psi_1 \rangle \langle \psi_1 |-|\psi_2 \rangle \langle \psi_2
|) +(p_3-\epsilon_1)(III +|\psi_1 \rangle \langle \psi_1 |-|\psi_2
\rangle \langle \psi_2 |)$$
$$+\epsilon_5(|\psi_5 \rangle \langle \psi_5 |+|\psi_6 \rangle \langle \psi_6 |)+\epsilon_7(|\psi_7 \rangle \langle \psi_7 |+|\psi_8 \rangle \langle \psi_8 |) ,$$
which is again a separable state. So in the second case, the
\emph{PPT} conditions are necessary and sufficient for
separability, too.
\section{Entanglement witnesses}
An entanglement witness acting on the Hilbert space
${\cal{H}}={\cal{H}}_{2}\otimes {\cal{H}}_{2}\otimes
{\cal{H}}_{2}$ is a Hermitian operator $\emph{W}=\emph{W}^{\dag}$,
that satisfies $Tr(\emph{W}\rho_s) \geq0$ for any separable state
$\rho_s$ in $ {\textbf{\emph{B}}}({\cal{H}})$ (Hilbert space of
bounded operators), and has at least one negative eigenvalue. If a
density matrix $\rho$ satisfies $Tr(\emph{W}\rho)<0$, then $\rho$
is an entangled state and we say that $\emph{W}$ detects
entanglement of the density matrix $\rho$. Note that in the
aforementioned definition of \emph{EW}s, we are not worry about
the kind of entanglement of the quantum state and we are rather
looking for \emph{EW}s which possess nonnegative expectation
values over all separable states despite of the fact that they
possess some negative eigenvalues. The existence of an \emph{EW}
for any entangled state is a direct consequence of Hahn-Banach
theorem \cite{Rudin} and the fact that the  subspace of separable
density operators is convex and closed \cite{{Jafarizadeh2}}.
Geometrically, \emph{EW}s can be viewed as hyper planes which
separate some entangled states from the set of separable states
and hyper plane indicated as a line corresponds to the state with
$Tr[W\rho]=0$.\par Based on the notion of partial transpose, the
\emph{EW}s are classified into two classes: decomposable
\emph{EW}s (d-\emph{EW}) and non-decomposable \emph{EW}s
(nd-\emph{EW}). An \emph{EW} $W$ is called decomposable if there
exist positive operators $ P, Q_{K} $ so that
\begin{equation}
 W=P+Q_1^{T_A}+Q_2^{T_B}+Q_3^{T_c} ,
\end{equation}
where $T_{K}$, $K=A,B,C$ denotes the partial transposition with
respect to subsystems $A,B$ and $C$, respectively. $W$ is called
non-decomposable if it can not be written in this form
\cite{Doherty}. Clearly a d-EW can not detect bound entangled
states (entangled states with positive partial transpose (PPT)
with respect to all subsystems) whereas there are some bound
entangled states which can be detected by a nd-EW.

Usually one is interested in finding optimal EWs $W$ which detect
entangled states in an optimal way. An EW $W$ is said to be
optimal, if for all positive operators $P$ and $\varepsilon>0$,
the new Hermitian operator
\begin{equation}\label{Wn2}
W'=(1+\varepsilon) W-\varepsilon P
\end{equation}
is not anymore an EW \cite{Lewenstein1}. Suppose that there is a
positive operator $P$ and $\epsilon\geq0$ such that
$W'=(1+\varepsilon) W-\varepsilon P $ is yet an EW
($Tr(W'\rho_s)\geq0$ for all separable states $\rho_s$). This
means that if $Tr(W\rho_s)=0$, then $Tr(P\rho_s)=0$, for all
separable states $\rho_s$ which indicates that, the operator $P$
is necessarily orthogonal to the kernel of $W$ denoted by
$Ker(W)$. By using the fact that every separable state is convex
combination of pure product states, one can take $\rho_s$ as a
pure product state $\ket{\psi}\langle\psi|$. Also, one can assume
that the positive operator $P$ is a pure projection operator,
since an arbitrary positive operator can be written as convex
combination of pure projection operators with positive
coefficients.
\subsection{EWs detecting bound MUB diagonal density matrices}
By employing tensor products of pauli operators relevant to
MUB-$(zzz)_G$ state of Table $I$ of the Appendix $A$, we introduce
the following linear three qubit EW \cite{Jafarizadeh1}
\begin{equation}\label{wit1}
 \\W=A_0III+A_1\ \ I
\sigma_{z}\sigma_{z}+A_2(\sigma_{x}\sigma_{x}\sigma_{x}
+\sigma_{x}\sigma_{y}\sigma_{y})+A_3(\sigma_{y}\sigma_{x}\sigma_{y}+\sigma_{y}\sigma_{y}\sigma_{x})
,
\end{equation}
where $A_0,A_2,A_3\geq0$ and $A_1$ can be negative or positive.
Now evaluating the trace of EW (\ref{wit1}) over a pure product
state,
$$\rho_s=| \alpha \rangle \langle \alpha |\otimes
    | \beta \rangle \langle \ \beta
    |\otimes | \gamma \rangle \langle \gamma|,$$
we get
$$Tr[W \rho_{s}]=A_0+A_1b_3c_3+A_2(a_1b_1c_1+a_1b_2c_2)+A_3(a_2b_1c_2+a_2b_2c_1)
 ,$$
where $a_ib_jc_k:=Tr[\sigma_i\sigma_j\sigma_k  \rho_s]$ for
$i,j,k=0,1,2,3$ with $a_0=b_0=c_0=1$. We parameterize points on
the unit sphere $S^2$ using traditional spherical coordinates, so
that $\theta$ and $\varphi$ stand for the angles of colatitude and
longitude, respectively $( \theta \in[0,\pi], \varphi
\in[0,2\pi])$. Thus, the points $a=(a_1,a_2,a_3)$,
$b=(b_1,b_2,b_3)$ and $ c=(c_1,c_2,c_3) $, can be uniquely
represented as the unit vectors with the following coordinates
$$a_1=\sin{\theta_1}\cos{\varphi_1},\;\ a_2=\sin{\theta_1}\sin{\varphi_1},\;\ a_3=\cos{\theta_1}$$
$$b_1=\sin{\theta_2}\cos{\varphi_2},\;\ b_2=\sin{\theta_2}\sin{\varphi_2},\;\ b_3=\cos{\theta_2}$$
$$c_1=\sin{\theta_3}\cos{\varphi_3},\;\ c_2=\sin{\theta_3}\sin{\varphi_3} ,\;\ c_3=\cos{\theta_3},$$
so that, we obtain $$Tr[W\rho_{s}]=
A_0+A_1\cos{\theta_3}\cos{\theta_2}+\sin{\theta_1}\sin{\theta_2}\sin{\theta_3}(A_2\cos{\varphi_1}\cos{(\varphi_2-\varphi_3)}+A_3\sin{\varphi_1}\sin{(\varphi_2+\varphi_3)})
.$$ By appropriate choice of the angles, one can minimize the
above expression. In fact, in the Appendix $B$, it has been proved
that by taking $A_0=\sqrt{A_2^2+A_3^2}$ and
$A_1=-\sqrt{A_2^2+A_3^2}$, the minimum value of $Tr[W\rho_{s}]$ is
attained to zero, i.e., we obtain
 $$min(Tr[W\rho_{s}])=0.$$
\par Consequently, EW (\ref{wit1}) takes the following form
\be\label{Wit3}W^{(\psi)}=\sqrt{A_2^2+A_3^2}(III-\ \ I
\sigma_{z}\sigma_{z}+\cos{\psi}(\sigma_{x}\sigma_{x}\sigma_{x}
+\sigma_{x}\sigma_{y}\sigma_{y})+\sin{\psi}(\sigma_{y}\sigma_{x}\sigma_{y}+\sigma_{y}\sigma_{y}\sigma_{x}))
,\ee where $$\cos{\psi}=\frac{A_2}{\sqrt{A_2^2+A_3^2}} ,\;\;\
\sin{\psi}=\frac{A_3}{\sqrt{A_2^2+A_3^2}}.$$ In the following, we
discuss the optimality of the obtained linear EW $W_{\psi}$.
\subsection{Optimality of the linear EW $W^{(\psi)}$}
According to the arguments about optimal EWs given in section $3$,
in order to prove the optimality of the EW $W^{(\psi)}$ given in
(\ref{Wit3}), it suffices to show that there exists no positive
operator $P$ such that $W':=(1+\varepsilon)W^{(\psi)}-\varepsilon
P$ be an EW, namely it must be proved that for any pure product
state $|\nu\rangle$ so that
$Tr(W^{(\psi)}|\nu\rangle\langle\nu|)=0$, there exists no positive
operator $P$ with the constraint $Tr(P|\nu\rangle\langle\nu|)=0$.
By considering a general three qubit pure product state as
\be\label{nu}\ket{\nu}=\frac{1}{\sqrt{2}}\left(\begin{array}{c}\cos{(\frac{\theta_1}{2})}\\
e^{i\varphi_1}\sin{(\frac{\theta_1}{2})}\end{array}\right)\otimes\left(\begin{array}{c}\cos{(\frac{\theta_2}{2})} \\
e^{i\varphi_2}\sin{(\frac{\theta_2}{2})}\end{array}\right)\otimes\left(\begin{array}{c}\cos{(\frac{\theta_3}{2})} \\
e^{i\varphi_3}\sin{(\frac{\theta_3}{2})}\end{array}\right) ,\ee
one can evaluate
$$Tr[W^{(\psi)}
\rho_s]=1-\cos{\theta_2}\cos{\theta_3}-
\sin{\theta_1}\sin{\theta_2}\sin{\theta_3}(\cos{\psi}\cos{\varphi_1}\cos{(\varphi_2-\varphi_3)}
+\sin{\psi}\sin{\varphi_1}\sin{(\varphi_2+\varphi_3})) ,$$ where
$\rho_s= \ket{\nu} \bra{\nu}.$ Now, it is easily seen that by
choosing the angles $\theta$ and $\varphi$ as follows
$$(1):\left\{\begin{array}{c}\cos{(\varphi_2-\varphi_3)}=1\\
\sin{(\varphi_2+\varphi_3)}=1\\
\end{array}\right.\Rightarrow \varphi_2=\varphi_3=\frac{\pi}{4}, \varphi_1=\psi ,
\theta_1=\frac{\pi}{2} , \theta_2=\theta_3 ,$$
$$(2):\left\{\begin{array}{c}\cos{(\varphi_2-\varphi_3)}=1\\
\sin{(\varphi_2+\varphi_3)}=-1\\
\end{array}\right.\Rightarrow \varphi_2=\varphi_3=-\frac{\pi}{4}  ,  \varphi_1=-\psi ,
\theta_1=\frac{\pi}{2} , \theta_2=\theta_3 ,$$
$$(3):\left\{\begin{array}{c}\cos{(\varphi_2-\varphi_3)}=-1\\
\sin{(\varphi_2+\varphi_3)}=-1\\
\end{array}\right.\Rightarrow \varphi_2=\frac{\pi}{4} , \varphi_3=-\frac{3\pi}{4}  ,  \varphi_1=\psi-\pi ,
\theta_1=\frac{\pi}{2} , \theta_2=\theta_3 ,$$
\be\label{ang}(4):\left\{\begin{array}{c}cos{(\varphi_2-\varphi_3)}=-1\\
\sin{(\varphi_2+\varphi_3)}=1\\
\end{array}\right.\Rightarrow \varphi_2=\frac{3\pi}{4}  \varphi_3=-\frac{\pi}{4}  ,  \varphi_1=\pi-\psi ,
\theta_1=\frac{\pi}{2} , \theta_2=\theta_3 ,\ee we obtain
$Tr[W^{(\psi)} \rho_s]=0$. Now, in order to prove that
$W^{(\psi)}$ is an optimal EW, we proceed as follows: Let $P$ be a
pure projection operator that one can subtract from $W^{(\psi)}$,
so that $(1+\epsilon)W^{(\psi)}-\epsilon P$ is an EW for some
$\epsilon>0$. From Eq. (\ref{Wit3}), one can easily see that, any
pure state of the form
$\ket{\Psi}=\ket{\alpha}\ket{z_+z_+}+\ket{\beta}\ket{z_-z_-}$,
(where, $\ket{\alpha}$ and $\ket{\beta}$ are arbitrary states)
belongs to the $Ker(W^{(\psi)})$, i.e., we have
$Tr[W^{(\psi)}\ket{\Psi}\langle\Psi|]=0$. Then, due to the fact
that, the pure projection operator $P$ must be orthogonal to
$Ker(W^{(\psi)})$ (and so orthogonal to $\ket{\Psi}\langle\Psi|$),
we have $P=\ket{\Phi}\langle \Phi|$ with
\begin{equation}\label{phi}
\ket{\Phi}=\ket{\alpha}\ket{ z_+z_-}+\ket{\beta}\ket{ z_-z_+}
\end{equation}  Now, the pure projection operator defined as above, must
be orthogonal to pure product states $\ket{\nu^{(i)}}$,
$i=1,2,3,4$ obtained by substituting the angles given by
(\ref{ang}) in (\ref{nu}), since these states belong to $Ker
(W^{(\psi)})$. But, this is possible only if
$\alpha_1,\alpha_2,\beta_1$ and $\beta_2$ satisfy the following
equations:
$$\bra{\nu_1}\Phi\rangle=(\alpha_1+\alpha_2
e^{-i\psi})+(\beta_1+\beta_2 e^{-i\psi})=0,$$
$$\bra{\nu_2}\Phi\rangle=(\alpha_1+\alpha_2
e^{+i\psi})+(\beta_1+\beta_2e^{+i\psi})=0,$$
$$\bra{\nu_3}\Phi\rangle=-(\alpha_1-\alpha_2e^{-i\psi})+(\beta_1-\beta_2e^{-i\psi})=0,$$
$$\bra{\nu_4}\Phi\rangle=-(\alpha_1-\alpha_2e^{+i\psi})+(\beta_1-\beta_2e^{+i\psi})=0.$$
Above equations imply that for$$\psi\neq0,\pi$$ we
have$$\alpha_1=\alpha_2=\beta_1=\beta_2=0.$$ Therefore, there is
no positive operator $P$ to subtract from $W^{(\psi)}$.\par
 In general, linear optimal EWs can be written as
 \begin{equation}\label{wit2}
W^{(\psi)}_{\pm i,\pm (j,k),(l,m)}=III\pm O_i+\cos{\psi}(O_j\pm
O_k)+\sin{\psi}(O_l\pm O_m),
\end{equation}
where $i=1,2,3$ while the indices $j \neq k \neq l \neq m$ take
values between $4,5,6,7$.\par The observables $O_i$ for $ i=1,2,3$
and $O_j$ for $j=4,5,6,7$ are defined as
$$O_1=I \sigma_{z}\sigma_{z},\;\ O_2=\sigma_{z} I \sigma_{z},\;\ O_3=\sigma_{z}\sigma_{z}
I,$$
$$O_4=\sigma_{x}\sigma_{x}\sigma_{x},\;\  O_5=\sigma_{x}\sigma_{y}\sigma_{y},\;\ O_6=\sigma_{y}\sigma_{x}\sigma_{y},\
O_7=\sigma_{y}\sigma_{y}\sigma_{x}.$$
\section{Non-linear optimal EWs}
Actually with a given entangled density matrix, one can associate
a non-linear EW, simply by defining a non-linear functional, so
that it is nonnegative valued over all separable density matrices,
but it is negative valued over the density matrix. In other words,
we optimize $Tr[W^{(\psi)}_{\pm i,\pm (j,k),(l,m)} \rho]$ where,
$W^{(\psi)}_{\pm i,\pm (j,k),(l,m)}$ are the linear optimal EWs
given by (\ref{wit2}) and $\rho$ is the
\emph{MUB}-$(\emph{zzz})_G$ diagonal density matrix given by
(\ref{bell ghz}). Then, one can easily get
$$Tr[W^{(\psi)}_{\pm i,\pm (j,k),(l,m)} \rho]=(1\pm r_i)+(r_j+r_k)\cos{\psi}+(r_l+r_m)\sin{\psi},$$
which indicates that, by appropriate choice of the parameter
$\psi$ as a functional of $\rho$, one can obtain a non-linear
function of the parameters of $\rho$ which is nonnegative over all
separable states. To this aim, we define
$$\cos{\theta}=\frac{r_j+r_k}{\sqrt{(r_j+r_k)^2+(r_l+r_m)^2}},$$
then $Tr[ W^{(\psi)}_{\pm i,\pm (j,k),(l,m)} \rho]$ can be written
as $$ Tr[W^{(\psi)}_{\pm i,\pm (j,k),(l,m)} \rho]=1\pm
r_i+\sqrt{(r_j+r_k)^2+(r_l+r_m)^2}cos{(\psi-\theta)}.$$ Now, by
choosing $(\psi-\theta)=\pi$, we obtain \be\label{result}Tr[W_{\pm
i,\pm (j,k),(l,m)} \rho]=1\pm
r_i-\sqrt{(r_j+r_k)^2+(r_l+r_m)^2}.\ee The above expression is the
required non-linear function in terms of the parameters of $\rho$
and it is definitely nonnegative valued function of separable
states, hence it is the non-linear optimal EW associated with
$\rho$ (since it is obtained from optimal linear EWs).
\subsection{Non-linear EWs as an envelop of family of linear EWs}
As the parameter $\psi$ of linear EW's $W^{(\psi)}_{\pm i,\pm
(j,k),(l,m)}$ varies, the envelope of hyper planes defined by
\be\label{wit4} Tr[W^{(\psi)}_{\pm i,\pm (j,k),(l,m)}\rho_s]=0,
\ee namely their intersections, define the boundary of PPT bound
entangled states that can be detected by the linear EWs.
Obviously, the envelope of these curves can be obtained simply by
eliminating the parameter $\psi$ from the Eq.(\ref{wit4}). To this
aim, we need to determine $\cos{\psi}$ \ and $\sin{\psi}$ by
solving above equation together with the equation that can be
obtained by taking its derivative with respect to $\psi$ equal to
zero, i.e., we consider
$$\left\{\begin{array}{c}Tr[W^{(\psi)}_{\pm i,+(4,5),(6,7)}\rho]=(1\pm
r_i)+(r_4+r_5)\cos{\psi}+(r_6+r_7)\sin{\psi}=0,\\
\hspace{-1cm}\frac{d}{d\psi}(Tr[W^{(\psi)}_{\pm
i,+(4,5),(6,7)}\rho])
=-\sin{\psi}(r_4+r_5)+\cos{\psi}(r_6+r_7)=0.\\
\end{array}\right.$$
By solving the above equations one can obtain
$$\cos{\psi}=-\frac{(r_j+r_k)(1\pm r_i)}{(r_j+r_k)^{2}+(r_l+r_m)^{2}},$$
$$\sin{\psi}=-\frac{(r_l+r_m)(1\pm r_i)}{(r_j+r_k)^{2}+(r_l+r_m)^{2}}.$$
Now, using the identity  $\cos^2{\psi}+\sin^2{\psi}=1$ we obtain
the required envelope of curves defined by the following equations
$$(1\pm r_i)^2=(r_j+r_k)^2+(r_l+r_m)^2 .$$
\section{Bound entangled MUB diagonal density matrices}
In this section, we consider three main categories of bound
entangled states according to equations (\ref{r1}) and (\ref{r2})
given in the Appendix $B$. In these categories, the relations
$|(r_j \pm r_k)|\leq (1\pm r_i) ,$ are always satisfied since if
we consider for example the inequality
$$|(r_4 - r_7)|>(1+ r_1), \ \ $$ then we conclude the inequality
$p_2 +p_4 < 0 $ which is clearly impossible.
\subsection{First  category }
The first interesting family of three qubit bound entangled states
is introduced for the choices of the parameters $r_i$ so that:
$$1\pm r_1=r_5\pm r_6,1\pm r_1=r_4\mp r_7,1\pm r_2=r_5\pm r_7,$$ \be\label{category1}1\pm r_2=r_4\mp r_6,1\pm r_3=r_6\pm r_7,1\pm r_3=r_4\mp r_5 ,\ee
for example, if we consider $1+r_1=r_4-r_7$, then we obtain $
p_2=p_4=0$. Then, the PPT conditions (\ref{p1})-(\ref{p6}) lead to
$p_1=p_3$, and triangle inequalities for the cases $(p_1,p_5,p_6)$
and $(p_1,p_7,p_8)$ are established. This state can be detected by
non-linear EW $W_{+1,-(4,7),(5,6)}$, since by using the result
(\ref{result}), we have \be\label{result1} Tr[W_{+1,-(4,7),(5,6)}
\rho]=(1+r_1)-\sqrt{(1+r_1)^2+(r_5-r_6)^2},\ee which indicates
that for $r_5\neq r_6$, $Tr[{W_{+1,-(4,7),(5,6)}\rho}]<0$.

On the other hand, by imposing the condition $(r_5-r_6)=p_5 -p_6
-p_7 + p_8 =0$, the state $\rho$ will be separable, since we have
$$p_5 -p_6 -p_7 + p_8 =0\Rightarrow
p_5+p_8=p_6+p_7 .$$So, by using the relations (\ref{rp}), one can
see that if $r_2>0$ ($r_2=2(p_6-p_8)$ since we have $p_1=p_3$ and
$p_2=p_4=0$), then we get $p_6>p_8$ and therefore we have
$$\rho=(1-r_2)III+(r_1+r_2)I\sigma_z\sigma_z+r_2[(III-\sigma_z\sigma_zI)(III+\sigma_zI\sigma_z)]+r_4\sigma_x\sigma_x\sigma_x+r_5\sigma_x\sigma_y\sigma_y$$
with
$$r_4-r_7=4p_1$$$$r_1+r_2=2(p_1-p_7-p_8)$$$$r_4-r_7-(r_1+r_2)=4p_1-2p_1+2p_7+2p_8=2(p_1+p_7+p_8)$$
$$1-r_2=1-2p_6+2p_8=2(p_1+p_7+p_8) $$
$$1-r_2-(r_1+r_2)+r_4-r_7=4(p_1+p_7+p_8)\leq1,$$
which is separable state.\\
If $r_2<0$ then we will have $p_8>p_6$ and
$$\rho=(1+r_2)III+(r_1-r_2)I\sigma_z\sigma_z-r_2[(III-\sigma_z\sigma_zI)(III+\sigma_zI\sigma_z)]+r_4\sigma_x\sigma_x\sigma_x+r_5\sigma_x\sigma_y\sigma_y$$
with$$r_1-r_2=2(p_1-p_5-p_6)\leq0$$$$r_4-r_7-(r_1-r_2)=2(p_1+p_5+p_6)\geq0$$$$1+r_2=2(p_1+p_5+p_6)\geq0$$
$$(1+r_2)+r_4-r_7-(r_1-r_2)=4(p_1+p_5+p_6)\leq1 .$$
For the above cases we had $ p_2=p_4=0$, so, this category
consists two vanishing non-paired ($p_2$ and $p_4$ belong to
different pairs) parameters. The other cases in (\ref{category1})
can be discussed similarly.\par
\textbf{A special case }\\
As a special case, if we consider $\rho$ in (\ref{bell ghz}) with
the following parameters
$$ p_4=p_8=p_6=0, $$
$$ p_3=p_5=p_7=p, $$  we get $r_1=r_2=r_3$ and $r_5=r_6=r_7$. Then, $\rho$ can be written as
\be \label{eq0}\rho=\frac{1}{8}[ I I I + r_{1} (I
\sigma_{z}\sigma_{z}+ \sigma_{z} I \sigma_{z}+\sigma_{z}\sigma_{z}
I )+ r_{4} \sigma_{x}\sigma_{x}\sigma_{x}+
r_{5}(\sigma_{x}\sigma_{y}\sigma_{y}+  \sigma_{y}
\sigma_{x}\sigma_{y}+  \sigma_{y} \sigma_{y}\sigma_{x}) ].\ee
Concerning the following PPT and normalization conditions
$$ p_1+p_2-p_3\geq0$$
 $$p_1-p_2+p_3\geq0$$
$$-p_1+p_2+p_3\geq0$$
 $$\sum_{i=1}^{8}p_{i}=1 \Rightarrow p_1+p_2+3p_3=1$$
we construct the convex hull of the following boundary planes
$\\p_1+p_2-p_3=0$ $\\p_1+p_2+3p_3=1$ $\Rightarrow  4p_3=1\
p_3=\frac{1}{4}$ $\\p_1-p_2+p_3=0$ $\\p_1+p_2+3p_3=1$ $\Rightarrow
2p_1+4p_3=1$ $\\-p_1+p_2+p_3=0$ $\\p_1+p_2+3p_3=1$ $\Rightarrow
2p_2+4p_3=1$,\par which define a triangular bound entangled region
in $(p_1 , p_2 , p_3)$ space (as it is shown in Fig.4). With the
boundary defined by the following lines where the boundaries of
triangular region corresponding to the lines passing through the
points $(\frac{1}{4},0,\frac{1}{4} ) ,( 0,\frac{1}{4},\frac{1}{4}
)$ and $(\frac{1}{2} ,\frac{1}{2} ,0)$, where the states
corresponding to the sides defined by the thick line passing
through the points $(\frac{1}{4},0,\frac{1}{4} )$ and $
(\frac{1}{2} ,\frac{1}{2} ,0)$ are separable.
\subsection{Second category  }
Another interesting family for bound entangled states is obtained
by considering the following cases:
$$1\pm r_1=r_4+r_5,1\pm r_1=r_4-r_5,1\pm r_1=r_6+r_7,1\pm r_1=r_6-r_7,$$ $$1\pm r_1=r_4+r_6,1\pm r_1=r_4-r_6,1\pm r_1=r_5+r_7,1\pm r_1=r_5-r_7,$$
$$1\pm r_2=r_4+r_5,1\pm r_2=r_4-r_5,1\pm r_2=r_6+r_7,1\pm r_2=r_6-r_7,$$ $$1\pm r_2=r_4+r_7,1\pm r_2=r_4-r_7,1\pm r_2=r_5+r_6,1\pm r_2=r_5-r_6,$$
\be\label{category2}1\pm r_3=r_4+r_6,1\pm r_3=r_4-r_6,1\pm r_3=r_5+r_7,1\pm r_3=r_5-r_7,$$ $$1\pm r_3=r_4+r_7,1\pm r_3=r_4-r_7,1\pm r_3=r_5+r_6,1\pm r_3=r_5-r_6.\ee
 If we choose one case such as $1+r_1=r_4+r_6 $, we obtain
 $p_7=p_1+p_2+2p_4+p_8$. Then, by using PPT conditions we obtain
 \be \label{s1} p_1+p_2-p_7+p_8\geq0\Rightarrow p_4=0 ,\ee
 \be \label{s2} p_3-p_7+p_8\geq0\Rightarrow p_3=p_1+p_2 ,\ee and satisfy the triangle inequality for $(p_3,p_5,p_6)$ case.
So according to (\ref{s1}) and (\ref{s2}) we get
\be\label{r1}p_7=p_3+p_8.\ee Applying the EW $W_{+1,+(4,6),(5,7)}$
to the state (\ref{eq0}), we obtain
$$Tr[W \rho]=(1+r_1)-\sqrt{(1+r_1)^2+(r_5+r_7)^2}<0$$
the condition $(r_5+r_7)=-p_1+p_2+p_5-p_6=0$ corresponds to
separable state.
\par By using the normalization condition
$$\sum_{i=1}^{8}p_{i}=1 \Rightarrow 3p_3+p_5+p_6+2p_8=1$$
and $$-p_1+p_2+p_5-p_6=0\Rightarrow p_1+p_6=p_2+p_5 ,$$ we obtain
for $r_2>0$
$$\rho=(1-r_2)III+(r_1+r_2)I\sigma_z\sigma_z+r_2[(III+\sigma_zI\sigma_z)(III-\sigma_z\sigma_zI)]+r_4\sigma_x\sigma_x\sigma_x+r_6\sigma_y\sigma_x\sigma_y$$
$$r_4+r_6=4(p_1+p_3)=4p_3$$$$r_1+r_2=-4p_8$$$$r_4+r_6-(r_1+r_2)=4p_7$$$$1-r_2=4p_7$$
$$\Rightarrow (1-r_2)+(r_4+r_6)-(r_1+r_2)=8p_7$$
while for $r_2<0$, we get\\
$$\rho=(1+r_2)III+(r_1-r_2)I\sigma_z\sigma_z-r_2[(III+\sigma_z\sigma_zI)(III-\sigma_zI\sigma_z)]+r_4\sigma_x\sigma_x\sigma_x+r_6\sigma_y\sigma_x\sigma_y$$
$$r_1-r_2=2(p_3-p_5-p_6)\leq0$$$$r_4+r_6-(r_1-r_2)=2(p_3+p_5+p_6)\geq0$$$$1+r_2=2(1-2p_7)=2(p_3+p_5+p_6)\geq0$$
$$\Rightarrow (1+r_2)+(r_4+r_6)-(r_1-r_2)=4(p_3+p_5+p_6)\leq1 .$$
So, in this case according to (\ref{r1}), if we choose parameters
as $p_i=p_j+p_k$, ($p_i$ and $p_j $ are in the same pairs) and
$p_k$ belong to another pairs, then we obtain second category of
bound entangled states.\par The other family (\ref{category2}) can
be considered similarly.
\subsection{Third category  }
The last category is given by the following cases:
$$1\pm r_1=r_4\pm r_7,1\pm r_1=r_5\mp r_6,1\pm r_2=r_4\pm r_6,$$ \be\label{category3}1\pm r_2=r_5\mp r_7,1\pm r_3=r_4\pm
r_5, 1\pm r_3=r_6\mp r_7 .\ee \par If $1-r_1=r_4-r_7$ then
$p_1+p_3=p_2+p_4+p_5+p_6+p_7+p_8=\frac{1}{2}$   and
$r_5-r_6=2(p_5-p_6-p_7+p_8).$ The PPT conditions for this case
have been previously considered (section 2.4.2). Therefore, if $r_5-r_6\neq0$
this state will be bound entangled and can be detected by
$W_{-1,-(4,7),(5,6)}$; otherwise it is separable since we have
$$r_5-r_6=0\Rightarrow p_6+p_7=p_5+p_8$$ and so from the PPT conditions we get
$$p_5=p_7 , p_6=p_8 $$ after calculation of $r_i$'s, we obtain
$$\rho=2p_2(III+\sigma_z\sigma_z I)(III+ I\sigma_z\sigma_z )+2p_4(III+ \sigma_z\sigma_z I )(III- \sigma_z
I\sigma_z)$$ $$+(1-2(p_2+p_4))III+2p_5
\sigma_x\sigma_x\sigma_x-4p_6\sigma_x\sigma_y\sigma_y.$$ We know
that the first and second cases in the density matrix are
separable and for other cases
$$|(1-2(p_2+p_4))|+|2p_5|+|-4p_6|\leq1 .$$ So, if we choose two
parameters and add them as $p_i+p_j=\frac{1}{2}$ ,( $p_i,p_j $ are
in different pairs) then, the category consists the bound
entangled states.\par The other cases (\ref{category2}) can be
discussed similarly.\par
\section{ Numerical analysis of three-qubit bound MUB diagonal
entangled states } This section is devoted to some   numerical
studies of three-qubit bound MUB diagonal entangled states as
follows: The feasible regions in  $(p_1,p_2)$ and $(p_1,p_3)$
planes, defined by equations (\ref{eq1}) and (\ref{eq2}), are
supported numerically. Using the nonlinear EWs, about $2.7\%$ of
bound MUB-$(zzz)_G$ diagonal density matrices are detected
numerically.
 The numerical results are plotted in $(p_1,p_3,p_5)$ , $(p_2,p_4,p_8)$
, $(p_6,p_7)$ and $(p_1,p_3)$ , $(p_2,p_4)$ , $(p_5,p_6)$ ,
$(p_7,p_8)$ phase spaces and  the  bound density matrix
($p_1=0.043425,p_2=0.15308,p_3=0.016132,p_4=0.19387,p_5=0.059793,p_6=0.24806,p_7=0.18207,p_8=0.10357$)
 is shown   in Fig.5 and Fig.6, as a prototype of a bound MUB diagonal
density matrix.
\section{Conclusion}
The feasible region of PPT \emph{MUB}-$(zzz)_G$ diagonal density
matrices is determined, where it is a convex polyatope due to
linearity of PPT conditions. In order to detect three-qubit bound
MUB diagonal entangled states, some nonlinear optimal EWs are
manipulated, such that they form the  envelope of the boundary
hyper-planes defined by a family of optimal linear EWs. By using
these nonlinear EWs, the region of bound entangled states and
separable ones are determined analytically in some particular cases
three categories,  where the numerical analysis support them. The
results thus obtained in this paper indicate that, the  proposed
methods in this work, can be used in studying entanglement of more
general systems with linear PPT conditions such as  multiqubit MUB
diagonal systems.
.\\\\
 \vspace{1cm}\setcounter{section}{0}
 \setcounter{equation}{0}
 \renewcommand{\theequation}{A-\roman{equation}}
  {\Large{Appendix A}}\\
\textbf{I. Mutually unbiased basis}\\ Let \emph{V} be a
\emph{d}-dimensional Hilbert space with two orthonormal basis
$$\emph{B}_1=\{ |e_1\rangle ,|e_2\rangle,...,|e_d\rangle \} \;\;\;\ \mbox{and}$$
$$\emph{B}_2=\{ |f_1\rangle ,|f_2\rangle,...,|f_d\rangle  \} ,$$
where  $\ket{e_i}$ and $\ket{f_i}$ for $i=1,2,...,d$ belong to $
\mathbb{C}^{d}$ (the standard Hilbert space of dimension $d$
endowed with usual inner product denoted by $\langle\  |\
\rangle$).

The basis $\emph{B}_1$ and $\emph{B}_2$ are called mutually
unbiased if and only if
\begin{equation} |\langle e_i |f_j \rangle |=\frac{1}{\sqrt{d}}.
\end{equation}
 As an example, for a two-level system there is such a set of bases
that can be represented in terms of eigenvectors of the usual
Pauli matrices $\sigma_x , \sigma_y , \sigma_z $ as follows
$$\emph{B}_x=\{\frac{1}{\sqrt{2}}(\ket{0}+ \ket{1}),\frac{1}{\sqrt{2}}(\ket{0}- \ket{1})\} ,$$
$$\emph{B}_y=\{\frac{1}{\sqrt{2}}(\ket{0}+ i\ket{1}),\frac{1}{\sqrt{2}}(\ket{0}- i\ket{1})\} \;\ \mbox{and}$$
$$\emph{B}_z=\{\ket{0} , \ket{1}\}.$$
When \textit{d} is a prime or power of a prime, the maximum number
of such MUB's  is equal to $\textit{d}+1$, otherwise there is no
clear number of sets.

According to Refs. \cite{Lawrence,Romero} and Table $I$, in the
case of three qubits, we have nine sets of mutually unbiased bases
and corresponding maximally commuting sets of observables, where
we will refer to them as generalized Pauli matrices; each of these
sets consist of seven commuting observables. In the Table $I$, the
first three rows contains product common eigenvectors,
$(\emph{xyz})_\pi ,(\emph{yzx})_\pi $ and $(\emph{zxy})_\pi$
(subscript $\pi$ means product state). For example, eight states
for basis $(\emph{xyz})_\pi$ could be written as $\ket{n_x n_y
n_z}$ where $n_i=1$ and $n_i=0$ correspond to spin down and spin
up along the $i$th axis for $i=x,y,z$, respectively. These product
states are separable so we will not use them for construction of
\emph{EWs}. Other bases consist of six maximally entangled states
$(\emph{xxx})_{Gi},(\emph{yyy})_G,(\emph{zzz})_G,(\emph{xzy})_G,(\emph{yxz})_G$
and  $(\emph{zyx})_G$, here subscript $G$ denotes a family of
Greenberger-Horne-Zeilinger (\emph{GHZ}) states. For example,
eight states for basis $(zzz)_G$ and $(xxx)_{Gi}$ can be written
as \be \label{zzzG}(zzz)_G=\ket{n_z n_z n_z ,\pm}=(\ket{n_z n_z
n_z}\pm\ket{\bar{n_z} \bar{n_z} \bar{n_z}}) , \;\ n_z=0,1\ee \be
\label{xxxG}(xxx)_{Gi}=\ket{n_x n_x n_x ,\pm}=(\ket{n_x n_x
n_x}\pm i\ket{\bar{n_x} \bar{n_x} \bar{n_x}}),\;\ n_x=0,1\ee
 where labels bar show that if $n_z=0$ or 1,
then $\bar{n_z}=1$ or 0, respectively. In section 2, we have
considered only the state $(zzz)_G$; since the other cases can be
obtained from this state by local unitary operations.\\
\textbf{II. MUB sets for three qubit systems}\\
We know MUB can be constructed using a number of methods that
depend on the dimensionality of the space. These methods using for
different case such as dimension space is prime, a product of
primes, or a power of a prime, and if it is odd or even. We
confine our study to the case of three qubits, that is, to an
eight-dimensional Hilbert space, in this space there exist four
MUB formation, where denotes sets of MUBs where the basis vectors
are either separable , biseparable or entangled states. The four
MUB formations are $(2,3,4)$ (here $2$ means two separable states,
$3$ \ means there is three biseparable state, $4$ means four
maximally entangled states), $(0,9,0)$, $(3,0,6)$ and  $(1,6,2)$.
There are other MUB sets for three qubits but one can get them by
local transformations from the previous ones. In this paper we
work with table $(3,0,6)$, line (6) in order to introduce our EWs.
This table is given by
\begin{table}[htbp]
$$\begin{tabular}{|c|c|c|c|c|c|c|c|c|}
  \hline
  $1$&$(xyz)_\pi$&$\sigma_x II$ & $I \sigma_y I$ & $II \sigma_z$ & $\sigma_x\sigma_y\sigma_z $& $\sigma_x\sigma_yI $& $\sigma_xI\sigma_z $&$ I\sigma_y\sigma_z $\\
 \hline
   $2$&$(yzx)_\pi$&$\sigma_y II$ & $I \sigma_z I$ & $II \sigma_x$& $\sigma_y\sigma_z\sigma_x $& $\sigma_y\sigma_zI $ & $\sigma_yI\sigma_x $ & $I\sigma_z\sigma_x $ \\
  \hline
   $3$&$(zxy)_\pi$&$\sigma_zII $& $I\sigma_xI $& $II\sigma_y $ & $\sigma_z\sigma_x\sigma_y $& $\sigma_z\sigma_xI $ & $\sigma_zI\sigma_y $ & $I\sigma_x\sigma_y $ \\
  \hline
  $4$&$(xxx)_{Gi}$&$\sigma_y\sigma_z\sigma_z $ & $\sigma_z\sigma_y\sigma_z $ & $\sigma_z\sigma_z\sigma_y $ &$\sigma_y\sigma_y\sigma_y $ & $\sigma_x\sigma_xI $ & $\sigma_xI\sigma_x $ & $I\sigma_x\sigma_x $ \\
  \hline
  $5$&$(yyy)_G$&$\sigma_z\sigma_x\sigma_x $ & $\sigma_x\sigma_z\sigma_x $ & $\sigma_x\sigma_x\sigma_z $& $\sigma_z\sigma_z\sigma_z $ & $\sigma_y\sigma_yI $ & $\sigma_yI\sigma_y $ & $I\sigma_y\sigma_y $ \\
  \hline
$6$&$(zzz)_G$&$\sigma_x\sigma_y\sigma_y $ & $\sigma_y\sigma_x\sigma_y $ & $\sigma_y\sigma_y\sigma_x $& $\sigma_x\sigma_x\sigma_x $ & $\sigma_z\sigma_zI $ & $\sigma_zI\sigma_z $ &$I\sigma_z\sigma_z $ \\
  \hline
  $7$&$(xzy)_G$&$\sigma_z\sigma_x\sigma_z $ & $\sigma_y\sigma_x\sigma_x $ & $\sigma_y\sigma_y\sigma_z $ & $\sigma_z\sigma_y\sigma_x $ &$\sigma_x\sigma_zI $ & $\sigma_xI\sigma_y $ & $I\sigma_z\sigma_y $ \\
  \hline
  $8$&$(yxz)_G$&$\sigma_x\sigma_y\sigma_x $ & $\sigma_z\sigma_y\sigma_y $ & $\sigma_z\sigma_z\sigma_x $ & $\sigma_x\sigma_z\sigma_y $ &$\sigma_y\sigma_xI $ & $\sigma_yI\sigma_z $ & $I\sigma_x\sigma_z $ \\
  \hline
  $9$&$(zyx)_G$&$\sigma_y\sigma_z\sigma_y $ &$\sigma_x\sigma_z\sigma_z $ &$\sigma_x\sigma_x\sigma_y $ & $\sigma_y\sigma_x\sigma_z $ & $\sigma_z\sigma_yI $ & $\sigma_zI\sigma_x $ & $I\sigma_y\sigma_x $ \\
  \hline
\end{tabular}$$
\caption{ Nine sets of operators defining a (3,0,6) MUB .}
\end{table}

In this table three states
 \be \label{zzz}
(xxx)_{Gi},(yyy)_G  , (zzz)_G\ee
 can be reversibly converted into
each other by local unitary operations (permutation), (i.e. $
\sigma_y\rightarrow \sigma_z\rightarrow \sigma_x$), and e.g., if
we construct EW using $(zzz)_G$ then this EW can be converted by
another EW for $(xxx)_G$ state by applying the local unitary
transformation $\sigma_z\rightarrow \sigma_x$ .\par Another states
\be \label{xzy}(xzy)_G,(yxz)_G ,(zyx)_G\ee can be transformed into
each other by local unitary operations, e.g., for state $(xzy)_G$
we have \be \label{xzyG}(xzy)_G=\ket{n_x^1 n_z^2 n_y^3
,\pm}=(\ket{n_x^1 n_z^2 n_y^3}\pm\ket{\bar{n_x^1} \bar{n_z^2}
\bar{n_y^3}}) ,\ee we can convert the states (\ref{xzy}) into
other three maximally entangled states (\ref{zzz}) by the
following local operations
$$
\begin{array}{cc}
  U_{x\leftrightarrow z}=\frac{1}{\sqrt{2}}\left(\begin{array}{cc}
  1 & 1 \\
  1 & -1 \\
\end{array}
\right),  &\quad U_{y\leftrightarrow x}= \left(
\begin{array}{cc}
  e^{\frac{i\pi}{4}} & 0 \\
  0 & e^{-\frac{i\pi}{4}}\\
\end{array}
\right),  U_{y\leftrightarrow z}=\frac{1}{\sqrt{2}} \left(
\begin{array}{cc}
  1 & i \\
  i & 1\\
\end{array}
\right). \\
\end{array}
$$
where, we have applied the permutation  $x\leftrightarrow z$ to
the first qubit, $y\leftrightarrow x$ to the middle qubit and
$y\leftrightarrow z$ to the rightmost qubit.\par The above
discussion was used for table $(3,0,6)$, but if we choose table
$(2,3,4)$, we will obtain maximally entangled states $((xzy)_G
,(yyz)_G,(yxy)_G,(zyx)_{Gi})$ so that, according to the following
local operators and permutations we can convert the states
(\ref{xzy}) into other three maximally entangled states. \par
$(yyz)_G  \rightarrow  (yxy)_G $  with local operators
$U(1)_{z\leftrightarrow x} ,U(2)_{y\leftrightarrow x} ,
U(3)_{z\leftrightarrow y}$ .(the notation $U(i)$ means that $U$
acting on the (i)-th qubit)\par $(xzy)_G  \rightarrow  (zyx)_{Gi}
$  with permutation  ($\sigma_x \rightarrow \sigma_z \rightarrow
\sigma_y$) only for first and second qubits and local operators $
U(3)_{y\leftrightarrow x}$\par $(xzy)_G  \rightarrow  (yyz)_G $
with local operators  $U(1)_{x\leftrightarrow y}
,U(2)_{z\leftrightarrow y}$ and with permutation ($\sigma_y
\rightarrow \sigma_z \rightarrow \sigma_x$) for the rightmost
qubit.\\\\
 \vspace{1cm}\setcounter{section}{0}
 \setcounter{equation}{0}
 \renewcommand{\theequation}{B-\roman{equation}}
  {\Large{Appendix B}}\\
 The following cases are used for construction of EWs:
\be\label{r1}r_4+r_5=2(p_3-p_4+p_5-p_6),r_6+r_7=2(-p_1+p_2+p_7-p_8)\ee
$$r_4-r_5=2(p_1-p_2+p_7-p_8) , r_6-r_7=2(p_3-p_4-p_5+p_6)$$ $$
r_4+r_6=2(p_3-p_4+p_7-p_8) ,r_5+r_7=2(-p_1+p_2+p_5-p_6)$$
$$r_4-r_6=2(p_1-p_2+p_5-p_6) ,r_5-r_7=2(p_3-p_4-p_7+p_8)$$
$$r_4+r_7=2(p_5-p_6+p_7-p_8) , r_5+r_6=2(-p_1+p_2+p_3-p_4)$$
$$r_4-r_7=2(p_1-p_2+p_3-p_4) , r_5-r_6=2(p_5-p_6-p_7+p_8).$$

We choose one the following cases for our EW:
\be\label{r2}1+r_1=2(p_1+p_2+p_3+p_4),\
1-r_1=2(p_5+p_6+p_7+p_8)\ee
$$1+r_2=2(p_1+p_2+p_5+p_6),\ 1-r_2=2(p_3+p_4+p_7+p_8)$$
$$1+r_3=2(p_1+p_2+p_7+p_8),\ 1-r_3=2(p_3+p_4+p_5+p_6).$$
\newpage
\textbf{Proof for EWs detecting bound MUB-$(zzz)_G$ diagonal density matrices}\\
For more detail, we have
$$Tr[W
 \rho_{s}]= A_0+A_1\cos{\theta_3}\cos{\theta_2}+\sin{\theta_1}\sin{\theta_2}\sin{\theta_3}(A_2\cos{\varphi_1}\cos{(\varphi_2-\varphi_3)}+A_3\sin{\varphi_1}\sin{(\varphi_2+\varphi_3)}) .$$
Taking $\varphi_2=\varphi_3=\frac{\pi}{4}$ ,
 $\theta_1=\frac{\pi}{2}$ \\ and according to relation
$(-\sqrt{a^2+b^2}\leq a\sin{\theta}+b \cos{\theta}\leq
\sqrt{a^2+b^2} )$, we have
$$
 Tr[W
 \rho_{s}]=A_0+A_1\cos{\theta_3}\cos{\theta_2}\pm\sin{\theta_2}\sin{\theta_3}\sqrt{A_2^2+A_3^3} ,$$
 according to $\cos{\theta_3}$, we obtain
 $$ Tr[W
 \rho_{s}]= A_0\pm\sqrt{A_1^2\cos{\theta_2}^2+\sin{\theta_2}^2(A_2^2+A_3^2)} ,$$
 and from the condition $(A_2^2+A_3^2\geq A_1^2 , \theta_2=
 \frac{\pi}{2})$, one can get
 $$\Rightarrow Tr[W
 \rho_{s}]= A_0\pm\sqrt{A_2^2+A_3^2} ,$$
 so , if$$A_0=\sqrt{A_2^2+A_3^2} ,$$
 $$Tr[W \rho_s]\geq0 ,$$
 $$W=\sqrt{A_2^2+A_3^2}III+A_1\ \ I
\sigma_{z}\sigma_{z}-A_2(\sigma_{x}\sigma_{x}\sigma_{x}
+\sigma_{x}\sigma_{y}\sigma_{y})+A_3(\sigma_{y}\sigma_{x}\sigma_{y}+\sigma_{y}\sigma_{y}\sigma_{x})
,$$ then  $A_1$ is an arbitrary number, therefore by taking
$A_1=-\sqrt{A_2^2+A_3^2}$, we get
$$W=\sqrt{A_2^2+A_3^2}(III-\ \ I
\sigma_{z}\sigma_{z}-\frac{A_2}{\sqrt{A_2^2+A_3^2}}(\sigma_{x}\sigma_{x}\sigma_{x}
+\sigma_{x}\sigma_{y}\sigma_{y})+\frac{A_3}{\sqrt{A_2^2+A_3^2}}(\sigma_{y}\sigma_{x}\sigma_{y}+\sigma_{y}\sigma_{y}\sigma_{x}))
.$$

\newpage
{\bf Figure Captions}

{\bf Figure-1:} Shows the feasible region in ( $p_1 , p_2 $) plane
defined by Eq.(\ref{eq1}).

{\bf Figure-2:} Shows the feasible region in ( $p_1 , p_3 $) plane
defined by Eq.(\ref{eq2}).

{\bf Figure-3:} The two colored triangles show the feasible region
obtained by the condition $p_1+p_3=\frac{1}{2}$, where the light
gray-colored triangle determines the region in ($p_3 , p_4 $)
plane and (according to the PPT conditions) the dark gray-colored
triangle indicates the region in ( $p_1 , p_2 $) plane. The dotted
line represents parametric equation (\ref{eq5}) which has been
drawn for two regions, i.e, from PPT conditions, dotted line in (
$p_3 , p_4 $) plane corresponds to the dotted line in ( $p_1 , p_2
$) plane.

{\bf Figure-4:} The gray-colored triangle shows the region of
bound entangled states in ($p_1 , p_2 , p_3$) phase space and
thick line shows the region of separable states.

{\bf Figure-5:}(a): Perspective of the region of bound entangled
states in $(p_1,p_3)$ plane obtained by numerical analysis. (b):
Perspective of the region of bound entangled states in $(p_2,p_4)$
plane obtained by numerical results. (c): Perspective of the region
of bound entangled states in $(p_5,p_6)$ plane obtained by numerical
analysis. (d): Perspective of the region of bound entangled states
in $(p_7,p_8)$ plane obtained by numerical analysis.

{\bf Figure-6:}.(a): Perspective of the region of bound entangled
states in $(p_1,p_3,p_5)$ phase space obtained by numerical
analysis. (b): Perspective of the region of bound entangled states
in $(p_2,p_4,p_8)$ phase space obtained by numerical analysis. (c):
Perspective of the region of bound entangled states in $(p_6,p_7)$
plane obtained by numerical analysis.


\newpage
\begin{thebibliography}{99}
\bibitem{Einstein} A. Einstein, B. Podolsky, and N. Rosen, Phys. Rev. 47, 777
~1935!; E. Schro¨dinger, Naturwissenschaften 23, 807 ~1935!; J. S.
Bell, Physics ~N.Y.! 1, 195 ~1964!.
\bibitem{Deutsch} D. Deutsch, Proc. R. Soc.
London, Ser. A 425, 73 ~1989!; P. Shor, SIAM J. Comput. 26, 1484
~1997!.
\bibitem{Ekert}A. Ekert, Phys. Rev. Lett. 67, 661 ~1991!.
\bibitem{Bennett}C. Bennett, G. Brassard, C. Crepeau, R. Jozsa, A. Peres, and W. K.
Wootters, Phys. Rev. Lett. 70, 1895 ~1993!.
\bibitem{Bouwmeester}D. Bouwmeester, J.-W. Pan, K. Mattle, M. Elbl, H. Weinfurter, and A.
Zeilinger, Nature ~London! 390, 575 ~1997!; D. Boschi, S. Brance, F.
de Martini, L. Hardy, and S. Popescu, Phys. Rev. Lett. 80, 1121
~1998!.
\bibitem{Wiesner}C. H. Bennett and S. J. Wiesner, Phys. Rev. Lett. 69, 2881 ~1992!.
\bibitem{Mattle}K. Mattle, H. Weinfurter, P. Kwiat, and A. Zeilinger, Phys. Rev.
Lett. 76, 4656 ~1996!.
\bibitem{Cleve}R. Cleve and H. Buhrman, Phys. Rev. A 56, 1201 ~1997!.
\bibitem{peres} A. Peres, Phys. Rev. Lett. {\bf 77}, 1413
(1996).
\bibitem{horodecki1} M. Horodecki, P. Horodecki, and R. Horodecki, Phys. Lett. A {\bf
223}, 1 (1996).
\bibitem{woronowicz} S. L. Woronowicz, Rep. on Math. Phys. {\bf  10}, 165
(1976).
\bibitem{horodecki2} P. Horodecki, Phys. Lett. A {\bf  232}, 333
(1997).
\bibitem{horodecki3} M. Horodecki, P. Horodecki, and R. Horodecki, Phys.
Rev. Lett. {\bf 80}, 5239 (1998).
\bibitem{terhal} B. M. Terhal, Phys. Lett. A{\bf  271}, 319
(2000).
\bibitem{jamiolkowski} A. Jamiolkowski, Rep. Math, Phys,
{\bf 3}, 275 (1972).
\bibitem{Wootters1} W. K. Wootters, Found. Phys. 16, 391 (1986).
\bibitem{Ivanovi}I. D. Ivanovi´c, J. Phys. A 14, 3241 (1981).
\bibitem{Lawrence} J. Lawrence, ¡C. Brukner, and A. Zeilinger, Phys. Rev. A 65, 032320
(2002).
\bibitem{Romero} J. L. Romero, G. Bj¨ork, A. B. Klimov, and L. L. S´anchez-Soto,e-print quant-ph/0508129v1
\bibitem{Rudin}W. Rudin, Functional Analysis,(McGraw-Hill, Singapore,1991).
\bibitem{Jafarizadeh2}M. A. Jafarizadeh, G. Najarbashi, and H. Habibian1,75, 052326 (2007)
\bibitem{Doherty} A. C. Doherty,P. A. Parrilo and F. M. Spedalieri, Phys. Rev. A{\bf 69}, 022308
(2004).
\bibitem{Lewenstein1}M. Lewenstein, B. Kraus, J. I. Cirac, and P. Horodecki, Phys. Rev.
A 62, 052310 (2000).
\bibitem{Jafarizadeh1}M. A. Jafarizadeh, M. Rezaee, and S. K. A. Seyed Yagoobi, Phys. Rev. A,72, 062106 (2005)
\bibitem{cirac} M. Lewenstein, B. Kraus, J. I. Cirac, P.
Horodecki, Phys. Rev. A{\bf  62}, 052310 (2000).


\end{thebibliography}
\end{document}